\documentstyle[pre,aps]{revtex}
\def\H{{\mathcal{H}}}
\begin{document}
\draft
\title{Generalized symmetric mutual information applied for the channel capacity}
\author{Takuya Yamano}
\address{Department of Applied Physics, Faculty of Science, Tokyo Institute of 
Technology, Oh-okayama, Meguro-ku, Tokyo,152-8551, Japan}
\maketitle
\begin{abstract}
The channel capacity for the binary symmetric channel is investigated based on the symmetrized definition of the 
mutual information, which is arising from an attempt of extension of information content based on the nonadditivity. 
The negative capacity can emerge as an avoidable consequence for the generalization of the concept of the information
 entropy when $q >1$. 
\end{abstract}
There has been a wide range of applications of the generalized (nonextensive) entropy by Tsallis\cite{88Tsallis}
to physical systems which exhibit power-low behaviour. 
Recently, Landsberg et.al. \cite{Landsberg} have focused on a possible constraint on the generalization of entropy 
concept and have compared channel capacities based on Shannon, Renyi, 
Tsallis, and another generalized entropy by applying to a binary symmetric channel (BSC) for a fixed 
error probability $e$ of communication.  It would be worth noting that the last generalized entropy introduced by 
Landsberg et al. , whose form takes as the Tsallis one divided by a factor $\sum_ip_i^q$ (we refer to it as the modified 
form of Tsallis entropy hereafter); the $p_i$ is the 
probability of the $i$-th state of the system considered and $q\in \mathcal{R}$, was presented independently 
by Rajagopal et al.\cite{RajAbe} from a consideration of form invariant structure of nonadditive entropy.

Motivated by these status, the present author has attempted to develop the source coding theorem based on a 
nonadditive information content\cite{Yamano} from the fact that the information entropy has the same 
form as the statistical mechanical entropy except for the Boltzmann factor in conventional definition. There we have chosen 
the nonadditive information content as $-\ln_qp(x)$ where $\ln_q x=(x^{1-q}-1)/(1-q)$. One should also 
notice that the normalized $q-$average\cite{98Tsallis} of $-\ln_qp(x)$ (information entropy) gives the modified form 
of Tsallis entropy\cite{Yamano}
 \begin{equation}
H_q(X) = \frac{\displaystyle{-\sum_{x\in \H}}p^q(x)\ln_qp(x)}
{\displaystyle{\sum_{x\in \H}}p^q(x)} = \frac{1-\displaystyle{\sum_{x\in \H}}
p^q(x)}{(q-1)\displaystyle{\sum_{x\in \H}}p^q(x)}\label{eqn:Hx},
\end{equation}
where $\H$ denotes alphabet to which a symbol $x$ belongs.  The concept of the mutual information can be regarded as 
the important ingredient to evaluate a communication channel in information theory\cite{Shan,ShanWea}. The mutual information is a measure of the amount of information that one random variable contains about another 
random variable. In both previous considerations\cite{Landsberg,Yamano},  there is a point that should be investigated 
about the definition of the mutual information. 
In \cite{Landsberg} they devised to take the average of conditional entropy $H(Y\mid X=x)$ 
using usual weighting $p(x)$ from the requisition that the capacity based on the any information entropy should be 
$0$ for $e=1/2$ as Shannon's case. However this selection possesses asymmetry of the mutual information in $X$ and $Y$ unless we 
choose the Shannon's $H(Y\mid X=x)$. The same applies to the case of \cite{Yamano}, where we followed the usual 
definition of the reduction in uncertainty due to another variable(the entropy subtracted by the conditional entropy).
In the nonadditive context, we may allow the mutual information to be not symmetric, however, if we decide to take the 
standpoint that any generalization of mutual information has to retain symmetry, we would have to accept the modification 
of the intuitively reasonable property of Shannon's capacity.\\

In this Letter we propose a new definition of the mutual information which holds symmetry in $X$ and $Y$ in the line of the 
previous work\cite{Yamano} and investigate the channel capacity of the BSC. 

Now we consider the mutual information indexed by $q$ as 
\begin{equation}
I_q(X;Y)\stackrel{\rm def}{=}H_q(X)-[1+(q-1)H_q(Y)]H_q(X\mid Y)
\end{equation}
 where $H_q(X\mid Y)$ is a nonadditive conditional entropy and is related with the nonadditive joint entropy$H_q(X, Y)$  as $H_q(X,Y)=H_q(Y)+H_q(X\mid Y)+(q-1)H_q(Y)H_q(X\mid Y)$\cite{RajAbe,Yamano}. Then we immediately see that 
the mutual information becomes symmetric in $X$ and $Y$,
\begin{equation}
I_q(X;Y)=I_q(Y;X)=H_q(X)+H_q(Y)-H_q(X,Y)\label{eqn:MI2}.
\end{equation}

In BSC $\H=\{0,1\}$ and the output ($y$) of the input code ($x$), $0$ and $1$ can be received as a $1$ and a $0$ respectively 
with the error probability $e$ due to  the external noise\cite{Cover}.  That is the conditional probabilities are given as $p(0\mid 1)=p(1\mid 0)=e$
, $p(0\mid 0)=p(1\mid 1)=1-e$ respectively. We write the input (output) probability which we observe a 
$0$ and a $1$ as $p_0$ and $p_1$ ($p_0^{\prime}$ and $p_1^{\prime}$) respectively. Then we have the joint probabilities $p_{xy}=p(y\mid x)p_x$; 
$p_{00}=(1-e)p_0$, $p_{01}=ep_0$, $p_{10}=ep_1$, and $p_{11}=(1-e)p_1$. \\
Using this the nonadditive joint entropy\cite{Yamano},
\begin{equation}
H_q(X,Y) = \frac{1-\sum_{x,y}p_{xy}^q}
{(q-1)\sum_{x,y}p_{xy}^q}
\end{equation}
is calculated as $(1-1/(p_0^q+p_1^q))(e^q+(1-e)^q))/(q-1)$. 

We define a generalized channel capacity as follows
\begin{equation}
C_q\stackrel{\rm def}{=}\max_{p(x)} I_q(X;Y)\label{eqn:Cq}.
\end{equation}
Since the output probabilities $p_0^\prime$ and $p_1^\prime$ are given as 
\begin{equation}
p_0^\prime=p_{00}+p_{10}=p_0+(p_1-p_0)e, \quad p_1^\prime =p_{01}+p_{11}=p_1+(p_0-p_1)e
\end{equation}
respectively, the summation $\sum_ip(y_i)$ is calculated as $((1-2e)p_0+e)^q+((2e-1)p_0+1-e)^q$.
Then from Eq(\ref{eqn:MI2}) the mutual information is expressed as 
\begin{equation}
I_q(X;Y)=\frac{1}{q-1}\{ \frac{1}{(p_0^q+(1-p_0)^q)}\left( 1-\frac{1}{e^q+(1-e)^q}\right)+\frac{1}{f(p_0,e)}-1\}
\end{equation}
where we have put $f(p_0,e)=((1-2e)p_0+e)^q+((2e-1)p_0+1-e)^q$. 
The $I_q(X;Y)$ can be taken its maximum when $p_0=p_1=1/2$ for some ranges of $q$ and $e$. Then the capacity is written as 
\begin{equation}
C_q=\frac{1}{q-1}\left( 2^q-1-\frac{2^{q-1}}{e^q+(1-e)^q}\right).
\end{equation}
Fig.1 shows the channel capacity $C_q$ against $e$ for the case of Shannon and different values of $q$. 
The capacity does not become zero for $e=1/2$ except for Shannon. It exceeds the Shannon's capacity for the 
intermediate noise level ($q<1$) and is below Shannon when the channel is very noisy(or less noisy).  
We note that the capacity can be negative also for our new definition of the mutual information for some ranges of $q$ and $e$ and 
we can have the capacity above Shannon for extreme cases of $e$. However, the negative channel capacity has a difficulty to understand 
in the usual communication framework as pointed out in the definition of \cite{Landsberg}.
 \\
 Summarizing, we have presented a new possible definition of the mutual information in nonadditive context, which preserves 
 the symmetry in the input and the output. As an application of this, the behaviour of the channel capacity of BSC 
 has shown for different noise characteristics.  The generalized channel capacity can entail negativity for $q >1$ when we sacrifice the symmetry 
 of mutual information  maintaining the zero capacity for $e=1/2$ and vice versa. The necessities for investigating whether or not the capacity based on the statistics breaking the additivity can be valid for real communication channel remains unchanged.
\acknowledgements
The author would like to thank Prof. A.K. Rajagopal for some comments.

{\bf Figure Caption}\\
\begin{figure}
\caption{The channel capacity of BSC is plotted against the error probability for Shannon and some different values of $q$.}
\label{Fig.1}
\end{figure}

\begin{references}
\bibitem{88Tsallis} C. Tsallis, J. Stat. Phys. {\bf 52}, 479 (1988); E.M.F. Curado and C. 
                    Tsallis, J.Phys.A {\bf 24}, L69 (1991); Corrigenda: {\bf 24}, 3187 (1991) 
                    and {\bf 25}, 1019 (1992).
\bibitem{Landsberg} P.T.Landsberg and V. Vedral, Phys. Lett. A {\bf 247}, 211 (1998).
\bibitem{RajAbe} A.K. Rajagopal and S. Abe, Phys. Rev. Lett. {\bf 83}, 1711 (1999).
\bibitem{Yamano} T.Yamano, cond-mat/0010074, Phys. Rev. E, in press.
\bibitem{98Tsallis} C. Tsallis, R.S. Mendes and A.R. Plastino, Physica A {\bf 261}, 534 (1998).
\bibitem{Shan} C.E.Shannon, Bell Syst. Tech. J. {\bf 27}, 379 (1948); \textit{ibid} 623 (1948).
\bibitem{ShanWea} C.E. Shannon and W.Weaver, \textit{The Mathematical Theory of Communication} 
                  (University of Illinois Press, Urbana,1963).
\bibitem{Cover} T.M. Cover and J.A. Thomas, \textit{Elements of Information Theory} 
                (Wiley,New York,1991).
\end{references}
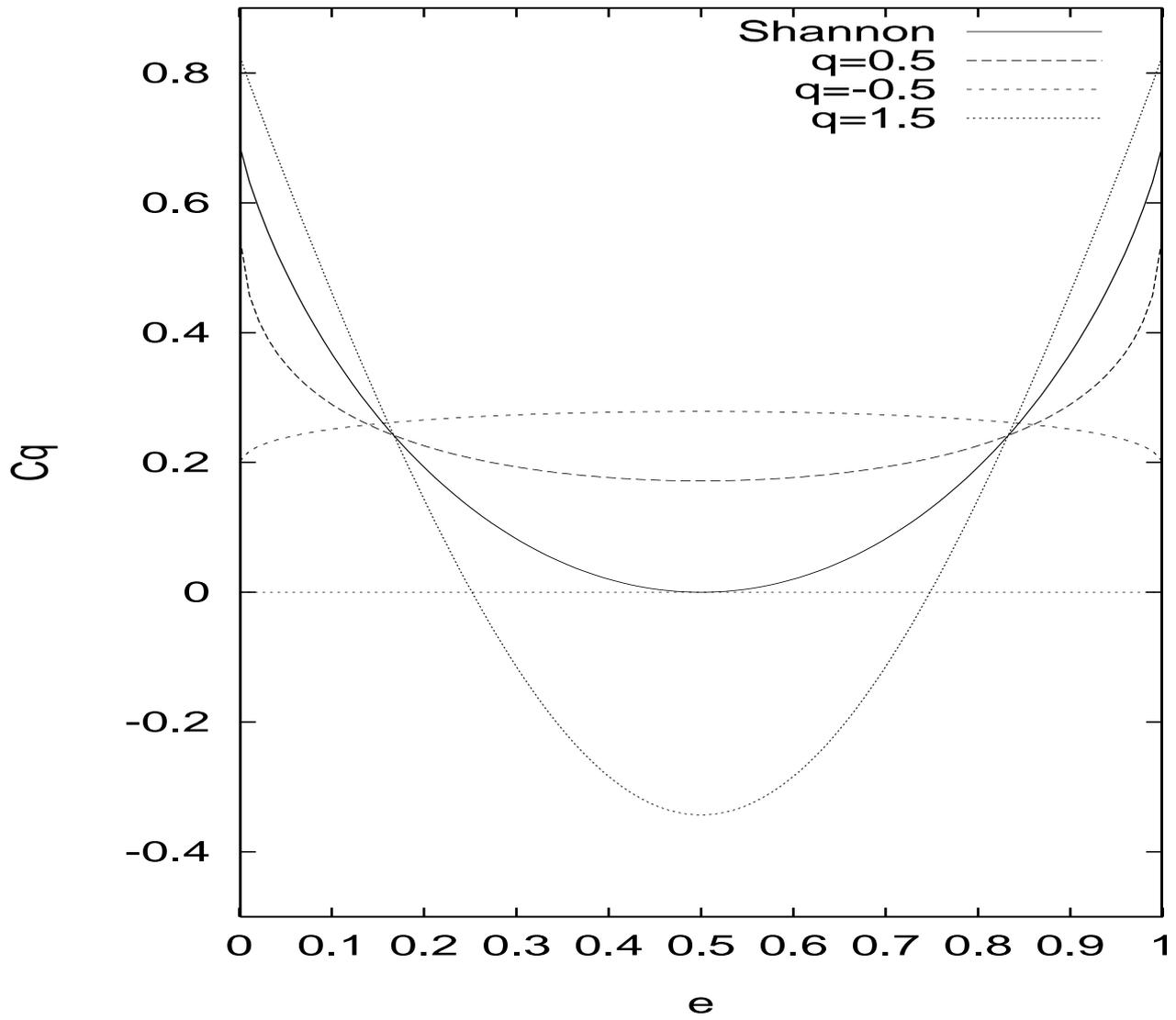
\end{document}